\documentclass[12pt,a4paper]{article}
\usepackage{amsmath}
\usepackage{amsfonts}
\usepackage{amssymb}
\usepackage{amsthm}
\usepackage{authblk}
\usepackage{placeins}
\usepackage{changepage}
\usepackage{graphicx}
\usepackage{chemformula}
\usepackage{xspace}
\usepackage{chicago}
\PassOptionsToPackage{hyphens}{url}\usepackage{hyperref}
\usepackage{booktabs}
\usepackage{tcolorbox}
\usepackage{setspace}
\newcommand{\gr}{``Green"\xspace}

\newcommand{\xca}{XCE\xspace}

\newcommand{\cote}{\ch{CO2}e\xspace}

\newcommand{\perm}{\text{perm}}
\newcommand{\Tperm}{\ensuremath{T_\perm}}

\theoremstyle{definition}
\newtheorem{manifesto}{Point}
\newtheorem{manifesto2}{Implication}

\begin{document}
\title{Transparency principle for carbon emissions drives sustainable finance} 

\author[1]{Chris Kenyon\thanks{Contact: c.kenyon@ucl.ac.uk}}
\author[2]{Mourad Berrahoui\thanks{Contact: ws802254@student.reading.ac.uk}}
\author[1,3]{Andrea Macrina\thanks{Contact: a.macrina@ucl.ac.uk}}
\affil[1]{Department of Mathematics, University College London}
\affil[2]{Henley Business School, University of Reading}
\affil[3]{African Institute of Financial Markets \& Risk Management University of Cape Town, South Africa}
\date{2022-02-15}
\maketitle

\newpage
\begin{abstract}
Alignment of financial market incentives and carbon emissions disincentives is key to limiting global warming \cite{bednar2021operationalizing}.  Regulators and standards bodies have made a start by requiring some carbon-related disclosures \cite{boe2021biennial} and proposing others \cite{anderson2019ifrs,pacf2020accounting,fsb2017crfd}.  Here we go further and propose a Carbon Equivalence Principle: all financial products shall contain a description of the equivalent carbon flows from greenhouse gases that the products enable, as well as their existing description in terms of cash flows.  This description of the carbon flows enabled by the project shall be compatible with existing bank systems that track cashflows so that carbon flows have equal standing to cash flows.  We demonstrate that this transparency alone can align incentives by applying it to project finance examples for power generation \cite{us2020capital,us2021capital} and by following through the financial analysis.  
The financial requirements to offset costs of carbon flows enabled in the future \cite{bertram2020ngfs} radically change project costs, and risk that assets become stranded, thus further increasing costs. This observation holds whichever partner in the project bears the enabled-carbon costs. Mitigating these risks requires project re-structuring to include negative emissions technologies \cite{fuss2018negative}.  We also consider that sequestered carbon needs to remain sequestered permanently, e.g., for at least one hundred years.  We introduce mixed financial-physical solutions to minimise this permanence cost, and price to them.  This complements previous insurance-based proposals \cite{marland2001accounting,coleman2018forest} with lesser scope. 
For financial viability we introduce project designs that are financially net-zero, and as a consequence are carbon negative.  Thus we see that adoption of the Carbon Equivalence Principle for financial products aligns incentives, requires product redesign, and is simply good financial management driving sustainability.
\end{abstract}

\newpage
\tableofcontents
\section{Introduction}
Sustainability, focusing on climate change, is a key issue for financial market participants, especially for banks \cite{lbg2020esg,mufg2021sustain} and their regulators \cite{boe2021biennial,ehlers2021taxonomy}.  WWF International \cite{WWF}, among many other global organisations, emphasise the crucial role financial organisations and systems play in decarbonising human life and in enabling drastic reduction of adverse impact to our planet.

Climate change is a result of increases in atmospheric greenhouse gasses (GHG) \cite{ipcc2021ar6}, and we consider these GHG in terms of equivalent carbon dioxide emissions (\cote).  Carbon pricing instruments, i.e., taxes or emissions trading systems, to control \cote emissions are already in operation in many jurisdictions. \cite{santikarn2021state}\ We propose the Carbon Equivalence Principle (CEP) as a manifesto: the carbon effects of a financial product shall be included as a second termsheet linked to the product's existing termsheet, effectively adding a new currency that we call \xca.  The letter X stands for non-national currencies, like XAU for gold, C for carbon, and E for equivalent, to achieve the required three-letter standard for a currency code.  By carbon effects of a financial product we mean the carbon flows that the product enables.  For example a loan financing a power plant enables the carbon emissions of the power plan over the life of the loan.  Paying off the loan returns the liability for future carbon flows to the payer.  The CEP means that all carbon flows are visible in volume and timing, and secondly that visibility follows the contractual relationships.  We claim that the CEP is necessary for accurate accounting in all institutions that have carbon targets, and is aligned with the industry Partnership for Carbon Accounting Financials (PACF) \cite{pacf2020accounting}.  Existing labels like \gr do not support the arithmetic of the carbon involved, neither in quantity, nor in timing. We further claim that the CEP is necessary for financial management: without \xca flows, assessing risks associated with future carbon pricing to the banks' portfolio cannot be carried out with standard bank systems as required by good governance.  We demonstrate the effects of CEP by applying it to project financing.  We consider new-build electricity generation in the US using a standard set of non-renewable and renewable technologies \cite{us2020capital,us2021capital}.  Our results emphasize that the CEP is simply a requirement from the point of view of banks' self-interest, i.e., normal financial management capturing all relevant future cashflows, and implications of future cashflows.

 Adoption of the CEP emphasises the financial target of net-zero effect from including equivalent carbon effects of financial products which requires net-negative \cote.

By using the CEP in a termsheet format, the \xca flows are compatible with existing bank systems, which in turn improves upon non-compatible disclosure proposals  \cite{fsb2017crfd,anderson2019ifrs,pra2019ss319,pacf2020accounting}.  The CEP provides transparency, thus enabling incentive alignment of decisions on financial products and sustainability.  Compatibility with existing bank systems means that all standard banking mechanisms can also be applied to carbon flows.  In particular, this includes quantification of future potential costs to projects and to all banks that are involved.  

Examples in project finance demonstrate that the CEP is necessary for normal financial analysis with respect to existing carbon price scenarios from the Network for Greening the Financial Markets (NGFS)  \cite{bertram2020ngfs}.   It is necessary because the financing costs can be significantly affected, and as a consequence that the projects may need to be redesigned both financially and physically.  In summary, our financial analysis of project finance examples with the CEP show:
\begin{itemize}
\item Enabled carbon emissions can add substantial costs to projects.  This is significant whichever entity is paying for offsets because carbon emissions offset costs can increase the project risk, and hence the financing costs.
\item  Power plants can become stranded assets given the carbon price projections of increasing offset costs.
\item Recovery rates, following financial default, may be much lower than  the historic 75\% \cite{sandp2021recovery} because of the risk that power plants may become stranded assets.   This means that the financing cost may be multiples of the initial estimate.
\item Projects may need to be redesigned.   For example, a second project for carbon sequestration may need to be added  to have cheaper carbon offset costs.
\item The cheapest carbon sequestration, afforestation and reforestation,  may not provide permanent carbon sequestration, for example, for at least one hundred years.  Thus we provide mixed financial/physical methods to create carbon sequestration permanence, and to price this sequestration permanence where required.  Previous  insurance-based proposals \cite{marland2001accounting,coleman2018forest} had smaller scope.
\item Projects can aim for financial net-zero, i.e., zero added costs from carbon emissions.  If more carbon is sequestered than emitted then the negative carbon balance may be sold.  Combined regulatory and voluntary carbon markets do not yet exist, but some COP26 statements expect such developments.  
\end{itemize}

The CEP complements and extends the carbon removal obligations (CROs) \cite{bednar2021operationalizing} in several ways.  First, we target future carbon flows.  Second, \xca termsheets are created for all financial products.  Third, and most importantly, the \xca termsheets create transparency only --- unlike CROs, the termsheets are not an automatic obligation, i.e., they do not require action.  Once \xca flows are visible, the markets will decide whether to reward or penalize.  That is, we claim that transparency and normal financial management are sufficient to change project financing costs and project design.   We demonstrate typical financial calculations in our numerical examples below.   For project finance carbon offset price curves are needed going out for thirty to seventy years.  If we include carbon sequestration permanence then we need carbon sequestration and carbon offset costs going out a hundred years and more.
Current carbon markets are limited, e.g., ICE carbon futures (see \url{https://www.theice.com/market-data/indices/commodity-indices/carbon-futures}) only go out about a year.  However, there are carbon price projections from inter-government sources, e.g., NGFS \cite{bertram2020ngfs} with similar maturities.  Where we require extrapolation we use extrapolation that is constant in real terms, i.e., we only adjust for projected inflation.  Thus we use a scenario-based approach, rather than replication and no-arbitrage pricing based on liquid markets \cite{Carmona2012,Borovkov2010,Ji2018}, because such liquid markets  do not exist.

Our numerical examples demonstrate the application of the CEP to project finance.  As  a principle, the CEP has much broader implications, but we think it is most important to have an in-depth demonstration of its use as a template for further applications.

This paper makes three main contributions.  First, we introduce the CEP, which includes removing the label \gr from financial products as inadequate, and demonstrates that most financial products are ESG-related.  Second, we provide quantitative  financial analysis for a project finance set, which demonstrates the wealth of insights and implications for finance and project design deriving from the application of the CEP.  Third, our analyses demonstrate that the Net-Zero Banking Alliance's commitment for member banks (representing over 40\%\ of bank assets) of net-zero emissions by 2050 \cite{unep2022nzba}  understates the financial requirement on projects, which is financial net-zero.  In other words, to have portfolios whose risk is not increased by enabling carbon emissions, implying that projects are carbon net-negative.

\section{Carbon Equivalence Principle (CEP)}

We state the CEP  and then give the main implications.  The CEP framework enables incentive alignment between sustainability and normal financial management.   Following the three points of the CEP, we provide a set of direct implications that arise from the application of the CEP.

\begin{tcolorbox}
	\begin{manifesto}{Carbon Equivalence for Financial Products I.}  The carbon-equivalent emissions, and carbon-equivalent sequestration, enabled and caused by a financial product shall be linked to the financial product.  {\it Enabled} means that the financial product makes these carbon flows possible.  {\it Caused by} means that carbon emissions are required for the lifecycle of the financial product. 
	\end{manifesto}
	\begin{manifesto}{Carbon Equivalence for Financial Products II.}  All financial products shall include the enabled carbon-equivalent impact as part of their definition, complementing their existing financial termsheets.  These carbon termsheets shall be compatible with existing financial position-keeping systems.
	\end{manifesto}
	\begin{manifesto}{No ``Green'' label.}  The label ``Green'' for financial products is inadequate and is superseded by the CEP.  
	\end{manifesto}
\end{tcolorbox}

A financial position-keeping system is the system used for storing financial product termsheets in electronic form that, at least, enables life-cycle events to be tracked and acted upon.  Life-cycle events are, for example, payment or receipt of cashflows on specified dates.  These events can then give rise to changes in treasury cash positions.  Pricing and risk management systems can be incorporated into position-keeping systems, or can be separate to just obtain positions (trades) from the position-keeping systems.

The first point of the CEP is to make the carbon flows enabled by a financial product visible, e.g., project finance for a gas-fired power station.  The first point is also to make the carbon flows caused by a financial product visible, for example bitcoin \cite{sedlmeir2020recent} or high frequency trading.  Most financial products in the retail and corporate areas will enable carbon emission or sequestration.  In contrast most financial products in the speculative area will cause carbon emission from their operation.

The second point states how the visibly will work.  That is, by capturing the carbon flows in the same position keeping systems that are used for usual cashflows and commodity flows.  Once within usual systems then usual pricing and risk management can be applied --- as we demonstrate below in project finance examples.  Note that the CEP requires visibility, but it does not contain any obligations for actions.  The CEP enables normal pricing and risk management from the visibility that the CEP requires.  An institution can ignore the carbon flows if it chooses to do so.  It is up to the market, and the pricing and risk management departments of institutions involved in transactions to penalise or reward based on their analyses.  Our examples below provide an example template for financial analyses, and for the potential structuring and project re-design consequent on these financial analyses.

The third point is to remove an inadequate label, \gr.  \gr is a binary label so it is not adequate to represent a quantity with a negative and positive range, and with a time dimension.  Several studies have not been able to find any material price effect produced by the \gr label on corporate \gr bonds \cite{tang2020shareholders,flammer2021corporate}.  This may simply represent the lack of information carried by the label, whatever certification it is based on.  The point is not just to remove an inadequate label, but to put the emphasis on sustainability, where we focus on carbon-equivalent emissions.  This is where the pricing and risk impact is for financial products.  The binary \gr label is appropriate where the \gr-ness is binary, e.g., green hydrogen, and may be retained in those cases.

The \xca termsheet can either be shorthand, i.e., aggregated into a single summary flow, or standard---by which we mean that all flows timings and quantities are given explicitly.  For some uses, summarisation into a single flow may be sufficient, so we include it as a possibility where appropriate.

We now list the direct implications of the CEP.
\begin{tcolorbox}

	\begin{manifesto2}\emph{All financial products are linked to climate change.}  The carbon equivalence demonstrates that all financial products have environmental (climate) impact in as much as they enable increased or decreased atmospheric carbon.  For many existing products the CEP will be material.   Moreover the CEP shows that recent discussions of ESG-linked financial products \cite{isda2021esg} are somewhat missing the point: financial products already have ESG consequences.  Having an additional ESG-link in the original termsheet is incomplete, and can be missing the main effects of carbon.
	\end{manifesto2}
	\begin{manifesto2}\emph{Product Maturity.}  The maturity of a financial product depends on both the non-\xca and \xca flows, and lifecycle events.  For example, requiring permanent sequestration of carbon creates a very long-dated financial product for Negative Emissions Technologies.
	\end{manifesto2}
	\begin{manifesto2}\emph{Price Discovery for Future Carbon Flows.} Current liquid carbon markets only go out about a year.  The CEP puts the emphasis on market players to discover the future prices of carbon flows. All stakeholders benefit from the visibility of the carbon flows linked to financial products of all maturities.
	\end{manifesto2}
	\begin{manifesto2}\emph{Consistency, Fungibility  and Structuring.}
		Application of the CEP results in all financial products being measured and priced consistently with respect to carbon flows.  There will be external consistency from the market, and internal consistency from pricing groups within projects and financial institutions. Once the impact is considered consistently then accounting is  simple.  With consistency, there is also fungibility, so structuring is simpler, i.e., informed decisions about warehousing, transfer, and offsetting.
	\end{manifesto2}
\end{tcolorbox}

\subsection{CEP application: power generation projects}

Here we consider how the CEP can be applied in project finance for new electricity generation based on  \cite{us2020capital,us2021capital}.  Our objective is to quantify the spread  on an annuity with the same notional as the capital cost of the power plants, and with varying maturities to represent different financing choices.  We do not consider the financing cost of the plants themselves, but only the cost differences  from the potential cost of carbon offsets according to NGFS scenarios.   Carbon impact can be classified into different scopes, typically direct and others \cite{ghg2015corporate}.  We focus on direct and include carbon emissions from construction, operation, and de-construction.   

We do not state who pays for the carbon offsets, or indeed if these need to be purchased at all.  Instead we quantify the potential risk as the possible carbon offset costs expressed in terms of a  spread on an annuity.  We express the cost as a spread on an annuity, because a spread automatically scales for both the capital quantity and the financing lifetime.  This spread is a standard way in banks to assess costs.  Note that it is a spread per year and so it is sometimes called a running spread.

Here we create the linked financial and carbon termsheets for project finance examples, where the projects are utility scale electricity generation, and consider the financial implications for financing.  That is, we calculate the projected cost to offset the carbon emissions of the power plants for the lifetime of the financing.  The cost is expressed  as basis points (0.01\%=1basis point, or bp) for an annuity of the length of the financing with the notional equal to the capital cost.  Only carbon emissions within the lifetime of the financing are considered.  This is because when the bond is redeemed by the issuer, the responsibility for enabling future carbon emissions follows the money and returns to the issuer.  This is the first point of the CEP.

Below we go through the points of the financial analysis listed in the introduction.  These range from quantifying carbon prices impacts for the length of the financing, through the implications of assets becoming stranded, to the necessity of accompanying NET projects,  the costs of making carbon sequestration permanent, and to the possibility of financial net-zero.

\subsection{Example of a linked termsheet}

This financing  structure is typical for any power plant.  Funding is acquired at the start of the project planning and we is repaid at financing maturity.  Financing maturity can be considerably before the end of power generation and de-construction of the power plant .  Carbon emissions start on construction start and end on de-construction end.  The difference between renewable power and combustion power is whether there are carbon emissions in operation.

In this example we assume two years planning, three years construction, then 40 years of operation, and finally 1.5 years de-construction when the site is returned to its original state.  Financing is required from the start so that long-lead time parts can be ordered.  

\vskip2mm
Part 1: 20-year, fixed-rate USD bond financing coal project.  
\begin{itemize}
	\item Notional is USD 1 billion
	\item Cashflows are Modified Following.  Daycount Act/360
	\item Party A pays Party B notional on 4 Jan 2022
	\item Party B pays Party A notional on 4 Jan 2042
	\item Party B pays Party A $x$\%\ annually on 29 Dec., first payment 2022, last payment 2042. 
\end{itemize}
Part 2: \xca flows.
\begin{itemize}
	\item Notional is M megawatts of electricity from coal power plant, type K
	\item Years 2 to 5: positive \xca M times $y_\text{construct}$\%\ where $y_\text{construct}$ is the construction cost in \xca per megawatt
	\item Years 5 to 45: positive \xca M times $y_\text{run}$\%\ where $y_\text{run}$ is the running cost in \xca per megawatt
	\item Years 45 to 46.5: positive \xca M times $y_\text{de-construct}$\%\ where $y_\text{de-construct}$ is the de-construction cost in \xca per megawatt
\end{itemize}
Values for each of the $x$ and $y_*$ constants are in Table \ref{t:technology}.  The linked termsheet has all of the XCA flows.  This creates transparency, while the carbon liability, under the CEP, only returns to the project once the loan is paid off. If the project defaults in construction and is not continued, then there are no subsequent carbon flows.  If  default occurs in operation, then restructuring of the loan and continued operation are usual.  However, if carbon offset costs are sufficiently high that the asset is uneconomic, then there is no restructuring and it becomes a stranded asset that cannot operate.

\section{Methods}

To apply the CEP we need to consider the potential consequences that the CEP makes visible.  Here we describe three that are relevant for application of the CEP to power plan construction, and are also general considerations: stranded assets; carbon sequestration permanence; and whether it is possible to have financial net-zero, not just carbon net-zero.  After this section we give the numerical results.

\subsection{Stranded assets and financing costs}

Increasing carbon prices over time mean that power generation assets can become stranded, i.e., abandoned.  Abandonment decreases project recovery rates because these assume financial restructuring and continued operation.  This change in recovery rate means that the initial financing spread must be larger.  The structure of project finance is that there is a loan at the start that is used to create an asset.  Once the asset is operational, profits from running the asset are used to pay down the loan.  Typically, if there is a default in the operation phase, then the loan is restructured and paid over a longer period with smaller payments.  Hence, historically, project finance has had much higher recovery rates than corporate loans \cite{sandp2021recovery}, roughly 75\%\ versus 40\%.

Considering the carbon liabilities of a project, we see that if a loan needs to be restructured over a longer period, then the asset's revenues will be paying higher costs because of rising carbon prices.  This means that the cash available for paying off the financing loan from restructuring will be less.  Hence, the asset may simply be uneconomic and be shut down, or the loan may need to be repaid over a longer period.  Meanwhile the carbon prices are still going up.  This is one mechanism for a carbon emitting asset to become stranded.  Thus the recovery rate needs to include the possibility that the asset becomes stranded.

Project finance recovery rates are high, around 75\%, so small changes in the recovery rate translate into large changes in the required financing spread. 
 The financing spread is linked to the CDS spread, because a CDS can be used to hedge default risk.  The CDS spread measures expected loss.
 The standard approximation for the relationship between hazard rate $\lambda$ and the spread on a Credit Default Swap (CDS) from \cite{brigo2006interest}:
\begin{align}
\lambda =& \frac{\text{CDS spread}}{1 - \text{recovery}}
\end{align}
If recovery changes from 75\%\ to 50\%, then the associated CDS spread doubles, if the hazard rate is unchanged.  

\subsection{Carbon permanence cost for NET}

Negative emissions technologies give a cost to capture carbon from the atmosphere, but they do not guarantee that this sequestration is permanent.  We assume that the project defines as permanence as continued sequestration of the same amount of carbon for the next $T_\text{perm}$ years after sequestration.  That is the original sequestered carbon may be released back into the atmosphere, but it must be immediately replaced by the same amount of sequestered carbon.

We use two standard financial concepts to address the permanence of carbon capture: default-and-repurchase, and custody.    Previous authors \cite{marland2001accounting,subak2003replacing,coleman2018forest}  proposed using insurance to create permanence, but the word insurance implies that the insurance cost is a fraction of the original cost.  We show here that the cost of permanence is multiples of the original cost.  That is, over a typical timescale required for permanence, say one hundred years, default can be expected to occur multiple times.

NET carbon capture can be reversed, both by financial events and by physical events. Financial events are relevant where the power plant project buys offset certificates from a company that then defaults. The new owner of the assets supporting the offset certificate may release the carbon, e.g., cuts down a forest for lumber and redevelops the land.  Physical events such as forest fires can release carbon captured by afforestation and reforestation.   Roughly 1\%\ of US forests burn each year \cite{usda2019forest,congress2021fires}.   In soil capture of carbon, the soil use may change following the default of the owning company and then the carbon may be released.  

We consider that the are two types of carbon capture:
\begin{itemize}
\item Physically-permanent capture, i.e., the captured carbon is captured in a  physically permanent way, e.g., as a solid.  This leads to custody solutions for permanence of carbon sequestration.
\item Physically-non-permanent capture, e.g., reforestation or afforestation.  This leads to default and repurchase solutions for permanence of carbon sequestration.
\end{itemize}
Custody means that a separate company takes physical possession of the captured carbon, and the only task of this custody company is to keep the captured carbon safe.   The contract with the custody company is such that if the financial rating of the custody company falls below a threshold it must replace itself with another company that is above the threshold rating.  In financial markets some institutions specialize in custody of assets, e.g., collateral like bonds or investments like precious metals.  They could expand into physically-permanent carbon capture.

Default and repurchase means that the carbon capture company defaults and all carbon capture certificates associated with it  default triggering a recovery rate payment.  The power plant project must then repurchase the emissions offset from a different carbon capture supplier at the time of default of the previous carbon capture supplier.  If carbon prices are increasing steeply, then it may be better to buy the additional offsets before the  carbon capture company defaults, rather than at default.  However, for simplicity, we only consider re-purchase upon supplier default.

We use a simple model to calculate the cost of permanence based on a Poisson process where we assume that financial default includes physical default.  In the US \cite{usda2019forest,congress2021fires}, the physical event hazard rate is about 1\%\ per year.  If the company owning the forest and issuing certificates has a BBB credit rating, then this implies a financial hazard rate of roughly 2.5\%\ / (1-recovery) per year, so financial default can be more important than physical default.  Here we assume that the company issuing certificates guarantees them for the life of the company, or permanence requirement, whichever is shorter. 

We assume that the power plant project creates a separate company to sequester the carbon, for example via a special purpose vehicle (SPV).  This SPV sequesters carbon and sells the certificates to the market.  If the SPV defaults, then the power plant project must create another SPV to sequester the carbon again.  

We can calculate the cost of permanence from $t$ to \Tperm, for a technology with sequestration cost $f(u)$ at time $u\in[t,\Tperm]$ as follows. We assume that the cost of replacing a sequestration company is $(1 - \textrm{recovery})$. So, if no carbon is released, recovery of value to creditors is 100\%\ and there is no cost.  Usually, however, financial recovery is less than perfect so there is some cost.  We suppose that the cost of replacing the SPV is proportional to the technology cost at the time, i.e., $f(u)$. Moreover, we assume defaults can be modelled by a non-homogeneous Poisson process with deterministic hazard rate $\lambda(u)$.

We begin by conditioning on the number of defaults $k$ that occur in the sequestration period $(t,\Tperm)$ to calculate the expected cost, the present value $PV_\perm$, at time $t\in[0,\Tperm)$. We have,
\begin{equation*}
PV_\perm(f, t, \Tperm, \lambda) = \sum_k \, (\text{Probability of } k\text{ defaults)} \text{(cost of }k\text{ defaults)}.
\end{equation*}
This means that defaults of sequestration companies are independent, and each sequestration company is identical from the point of view of default.  The default hazard process is given by 
\begin{align*}
\Lambda(t,\Tperm) = \int_{s=t}^{s=\Tperm} \lambda(s) ds.
\end{align*}
The probability of $k$ events within the period $(t,\Tperm)$ is the Poisson probability $P_o(k,\Lambda(t,\Tperm))$.
So the expected cost of permanence is:
\begin{align*}
PV_\perm(f, t, \Tperm, \lambda) &= \sum_{k=0}^{\infty} \, (\text{Probability of } k\text{ defaults)} k \text{(cost of }1\text{ default)} \\
&= \sum_{k=0}^{\infty} P_o(k,\Lambda(t,\Tperm))\ k\ C(t,\Tperm,\lambda),
\end{align*} 
where $C(t,\Tperm,\lambda)$ is the cost of one default within the period $[t,\Tperm]$. 
From the properties of a Poisson process, we know that given that one event has occurred in $(t,\Tperm)$  the likelihood of the event occurring at $u\in(t,\Tperm)$ is $\lambda(u)/\Lambda(t,\Tperm)$. Now, if the default happens at time $u$, the default cost will be $f(u)(1-R)$ per unit of sequestration, where we use $R$ for the recovery percentage.  We discount that cost back to $t$, by using the funding cost of the project, with the discount funding bond $D_F(t,u)$, to give
\begin{align*}
C(t,\Tperm,\lambda) &= \int_{u=t}^{u=\Tperm} D_F(t,u) f(u) (1-R)  \frac{\lambda(u)}{\Lambda(t,\Tperm)} du.
\end{align*}
Putting this together gives:
\begin{align*}
&PV_\perm(f, t, \Tperm, \lambda)\nonumber\\
&= \sum_{k=0}^{\infty}P_o(k,\Lambda(t,\Tperm))\ k\ \int_{u=t}^{u=\Tperm} D_F(t,u) f(u) (1-R) \frac{\lambda(u)}{\Lambda(t,\Tperm)} du.
\end{align*}
For simplicity we assume that $\lambda(t)$ is constant (i.e., $\lambda$), and so we have
\begin{align}
&PV_\perm(f, t, \Tperm, \lambda)\nonumber\\
&= \sum_{k=0}^{\infty}P_o(k,\lambda(\Tperm-t))\ k\ \int_{u=t}^{u=\Tperm} D_F(t,u) f(u) (1-R) \frac{\lambda}{\lambda(\Tperm-t)} du  \nonumber \\
&=\lambda(\Tperm-t) \frac{1}{\Tperm-t}\int_{u=t}^{u=\Tperm} D_F(t,u) f(u) (1-R)  du \nonumber\\
&=\lambda (1-R) \int_{u=t}^{u=\Tperm} D_F(t,u) f(u)   du \label{e:perm}.
\end{align}
Equation \ref{e:perm} describes a permanence curve, i.e., the cost of permanence for carbon sequestration required at time $t$ until \Tperm. Permanence may be required up to project start plus, say, 100 years, or 100 years from the original carbon sequestration date.  In our calculations we assume the latter.

\subsection{Financial net-zero}

Future carbon prices under NGFS scenarios are more expensive than the cheapest NET, affoforestation and reforestation, even when permanence-adjusted.  Thus we can look for a financial net-zero solution where the power project builds NET capacity, and sells all of the capacity, so as to have zero effect from the emission and sequestration of carbon.  

Table \ref{t:fnz} shows the NET capacity in millions of tonnes (MT) of carbon captured per year, as required under each NGFS scenario to achieve financial net-zero for the lifetime of the financing.  That is, the NET are sold at the market price to pay for the cost of the carbon emissions.  NGFS price scenarios  already include the existence of NET capacity \cite{bednar2021operationalizing} in the market price of carbon.  The power project can be part of this capacity and profit from it sufficiently to be carbon negative and financially net-zero.  The reason for the difference between the NET cost and the carbon price is that limited capacity to build NET is assumed.  Thus financial net-zero depends on the project's ability to build NET capacity.  

\section{Numerical results}

The array of technologies considered is roughly half/half between non-renewable and renewable electricity generation methods.  Whilst the renewable methods are assumed to have no material carbon emissions in operation, their construction and de-construction are carbon emitting. 
\begin{figure}[th]
	\begin{center}
		\includegraphics[trim=0 0 0 0, clip, width=0.49\textwidth]{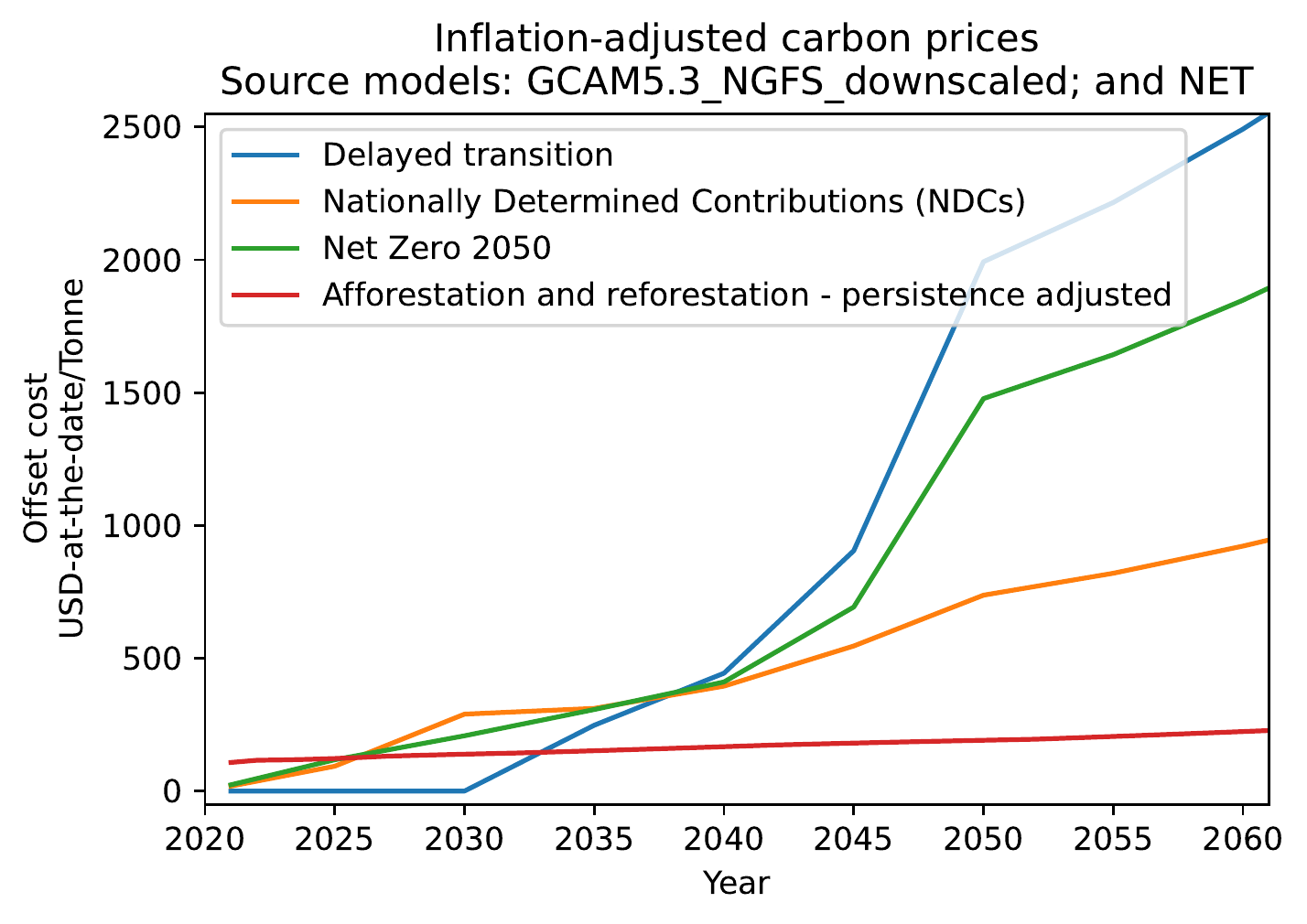}
		\includegraphics[trim=0 0 0 0, clip, width=0.49\textwidth]{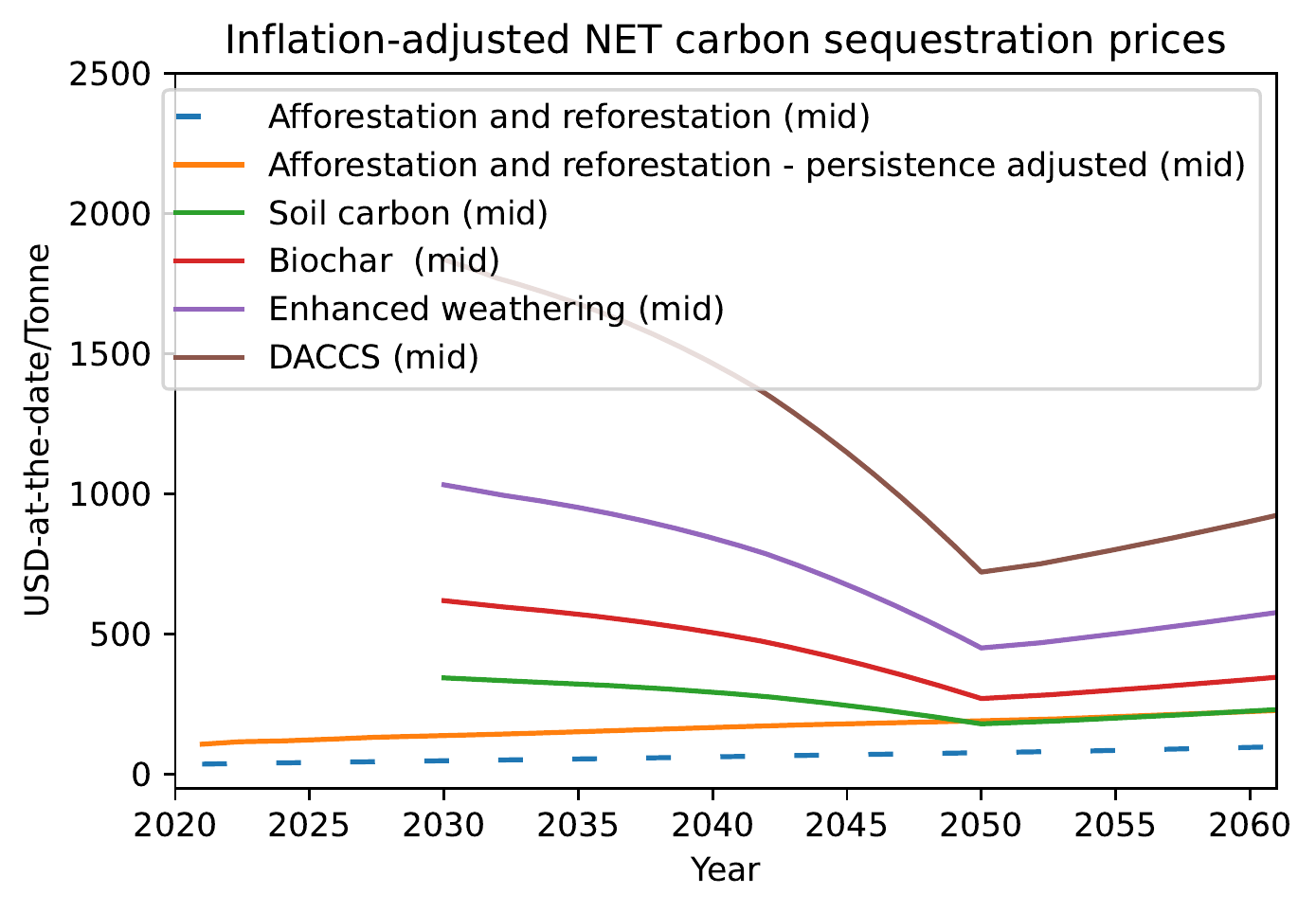}
	\end{center}
	\caption{[LEFT] Carbon price projections for USA from NGFS.  Source data (c) IIASA 2021, adapted for 2021 baseline and future dollars, i.e., inflation adjusted, from 2010 dollars.  These NGFS scenarios stop at 2050 and we have extrapolated using flat 2010 dollars, then inflation adjusted to extend.   Used in accordance with licensing \url{https://data.ene.iiasa.ac.at/ngfs/\#/license}. [RIGHT] projected prices for NET.  NET are assumed mature in 2050 and have an emergent technology multiplier before 2030, and are also inflation adjusted.  Data source and treatment described in the text.}
	\label{f:ngfs}
\end{figure}
Setup and calculation are given in the Methods section.  Figure \ref{f:ngfs} shows a graph of the NGFS inflation-adjusted prices under three scenarios: Delayed Transition; Nationally Determined Contributions (NDC); and Net-Zero 2050.  A fourth curve overlays the cheapest NET technology where the cost has been adjusted to price in 100-year permanence from date of sequestration.

\begin{figure}[th]
	\begin{center}
		\includegraphics[trim=0 0 0 0, clip, width=0.49\textwidth]{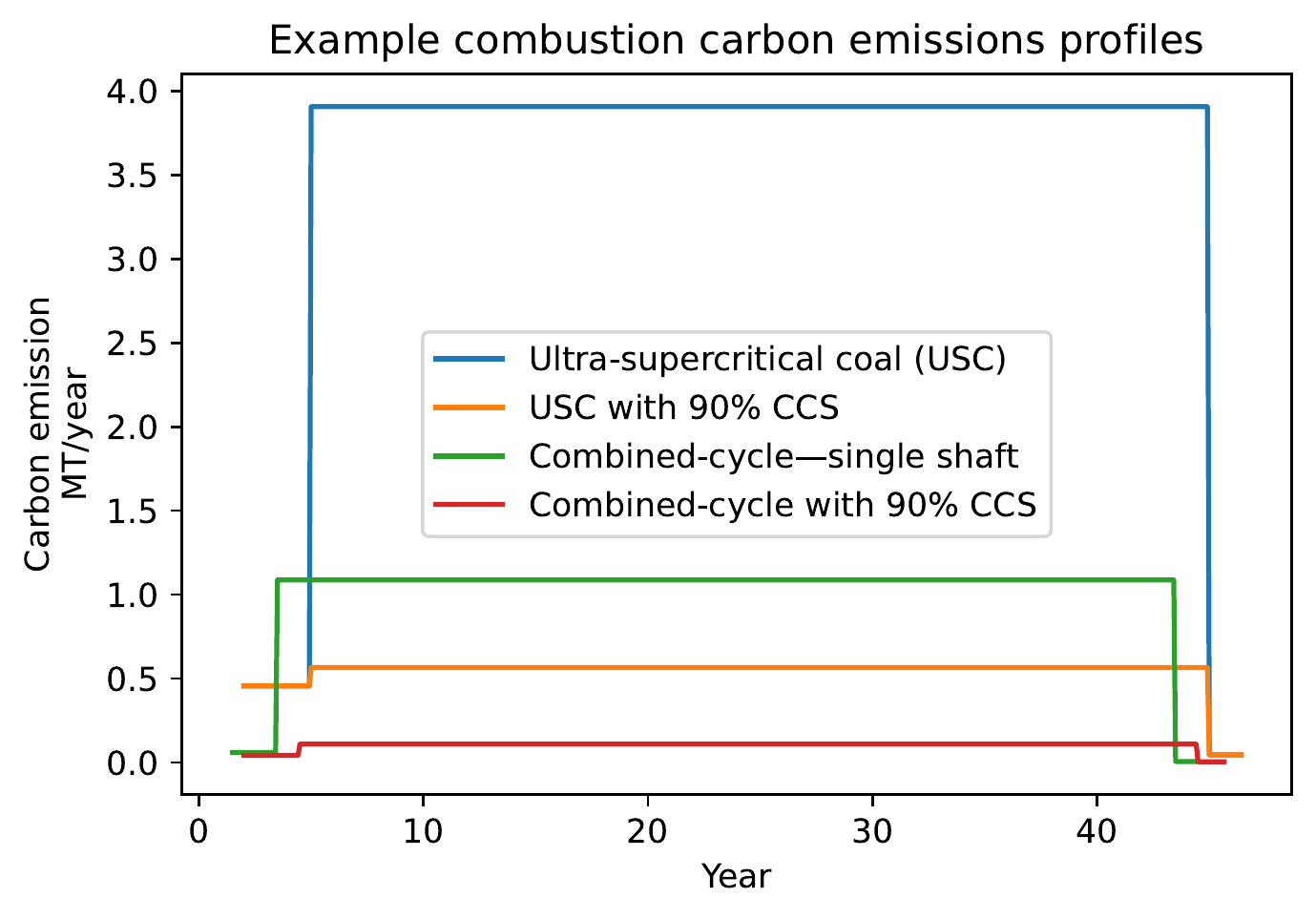}
		\includegraphics[trim=0 0 0 0, clip, width=0.49\textwidth]{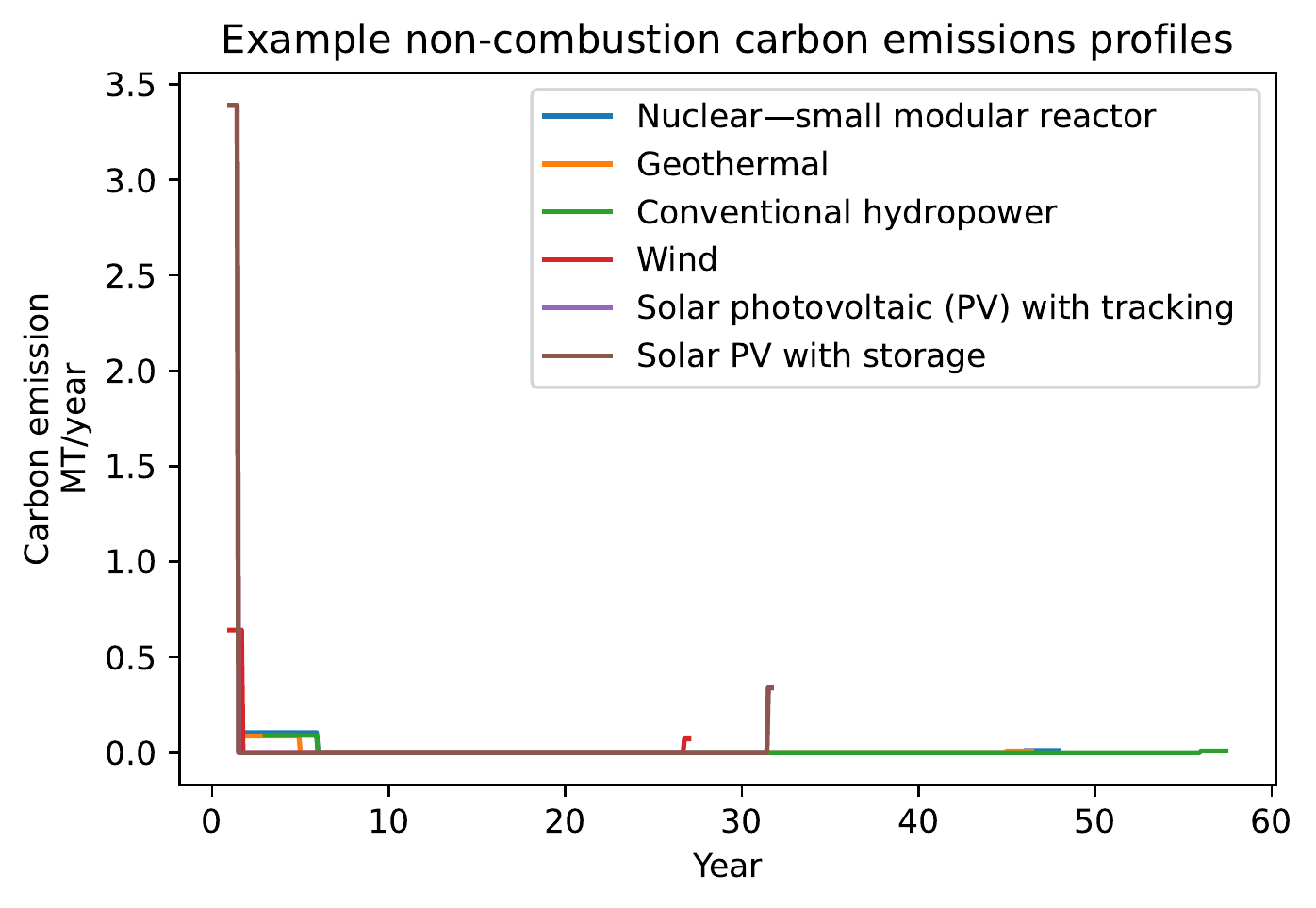}
	\end{center}
	\caption{Carbon emissions profiles for the cases considered in the text covering construction, operation, and deconstruction.  Total emissions are the areas under the curves.   [LEFT] Combustion power plants, with and without 90\%\ CCS. [RIGHT] Non-combustion (renewable) power plants, note that there can be significant emissions for construction.}
	\label{f:carbonProfiles}
\end{figure}

Figure \ref{f:carbonProfiles} shows the carbon emissions profiles for the power plants considered.  The combustion power plants have most emissions in operation, whereas the non-combustion power plants have most emission on construction, and a second peak for de-construction.

\begin{table}
	\vspace{0cm}
	\begin{adjustwidth}{-1.25cm}{-0cm}
	\begin{center}
		\begin{tabular}{llrrrr}
			\toprule
 Units are bps of annuity spread                         & Maturity Case &   10y &   20y &   30y &    All \\
Case no., Technology & NGFS scenario &       &       &       &        \\
\midrule
01, Ultra-supercritical coal (USC) & Delayed transition &    30 &  1779 &  5673 &  11766 \\
                          & Forest &  1030 &  1605 &  1893 &   2152 \\
                          & NDCs &  1647 &  3147 &  4539 &   6274 \\
                          & net-zero 2050 &  1340 &  2899 &  5503 &   9744 \\
03, USC with 90\% CCS & Delayed transition &     3 &   161 &   514 &   1070 \\
                          & Forest &   130 &   166 &   187 &    208 \\
                          & NDCs &   175 &   300 &   422 &    579 \\
                          & net-zero 2050 &   153 &   281 &   512 &    896 \\
08, Combined-cycle—single shaft & Delayed transition &    45 &  2611 &  8326 &  16574 \\
                          & Forest &  1871 &  2562 &  2936 &   3259 \\
                          & NDCs &  2723 &  4795 &  6795 &   9118 \\
                          & net-zero 2050 &  2323 &  4459 &  8232 &  13943 \\
09, Combined-cycle with 90\% CCS & Delayed transition &     2 &   126 &   403 &    824 \\
                          & Forest &    89 &   123 &   141 &    158 \\
                          & NDCs &   129 &   231 &   328 &    446 \\
                          & net-zero 2050 &   109 &   214 &   397 &    689 \\
\midrule
12, Nuclear—small modular reactor & Delayed transition &     0 &     0 &     0 &      2 \\
                          & Forest &    14 &     8 &     6 &      5 \\
                          & NDCs &    11 &     6 &     5 &      4 \\
                          & net-zero 2050 &    12 &     7 &     5 &      5 \\
15, Geothermal & Delayed transition &     0 &     0 &     0 &     28 \\
                          & Forest &   249 &   143 &   109 &     88 \\
                          & NDCs &   177 &   102 &    77 &     71 \\
                          & net-zero 2050 &   213 &   123 &    93 &     94 \\
17, Conventional hydropower & Delayed transition &     0 &     0 &     0 &      6 \\
                          & Forest &    61 &    35 &    27 &     20 \\
                          & NDCs &    56 &    32 &    24 &     20 \\
                          & net-zero 2050 &    61 &    35 &    27 &     24 \\
20, Wind & Delayed transition &     0 &     0 &    25 &     25 \\
                          & Forest &   232 &   133 &   111 &    111 \\
                          & NDCs &    87 &    50 &    51 &     51 \\
                          & net-zero 2050 &   110 &    63 &    70 &     70 \\
24, Solar photovoltaic (PV) with tracking & Delayed transition &     0 &     0 &     0 &    108 \\
                          & Forest &  1050 &   603 &   459 &    451 \\
                          & NDCs &   373 &   214 &   163 &    197 \\
                          & net-zero 2050 &   470 &   270 &   205 &    278 \\
25, Solar PV with storage & Delayed transition &     0 &     0 &     0 &     81 \\
                          & Forest &   786 &   451 &   343 &    337 \\
                          & NDCs &   279 &   160 &   122 &    147 \\
                          & net-zero 2050 &   352 &   202 &   154 &    208 \\
\bottomrule
\end{tabular}
		\caption{Annuity spread in basis points (bps) for different bond maturities to cover carbon emissions liabilities during the financing lifetime of the bond.  NDCs = Nationally Determined Contributions.}
		\label{t:spread}
		\end{center}
	\end{adjustwidth}
\end{table}

\begin{table}
	\vspace{0cm}
	\begin{adjustwidth}{-1.25cm}{-0cm}
	\begin{center}
		\begin{tabular}{llrrrr}
			\toprule
  Units are MT/year of NET capacity                        & Maturity Case &   10y &   20y &   30y &    All \\
Case no., Technology & NGFS scenario &       &       &       &        \\
\midrule
01, Ultra-supercritical coal (USC) & Delayed transition &    NA &  6.6 &  4.5 &  4.1 \\
                          & NDCs &   7.6 &  7.1 &  6.1 &  5.4 \\
                          & net-zero 2050 &  11.5 &  7.5 &  5.2 &  4.5 \\
03, USC with 90\% CCS & Delayed transition &    NA &  1.0 &  0.6 &  0.6 \\
                          & NDCs &   1.3 &  1.1 &  0.9 &  0.8 \\
                          & net-zero 2050 &   2.1 &  1.2 &  0.8 &  0.7 \\
08, Combined-cycle—single shaft & Delayed transition &    NA &  1.8 &  1.2 &  1.2 \\
                          & NDCs &   2.4 &  2.0 &  1.7 &  1.5 \\
                          & net-zero 2050 &   3.8 &  2.2 &  1.5 &  1.3 \\
09, Combined-cycle with 90\% CCS & Delayed transition &    NA &  0.2 &  0.1 &  0.1 \\
                          & NDCs &   0.2 &  0.2 &  0.2 &  0.2 \\
                          & net-zero 2050 &   0.4 &  0.2 &  0.1 &  0.1 \\
\midrule
12, Nuclear—small modular reactor & Delayed transition &    NA &  0.0 &  0.0 &  0.0 \\
                          & NDCs &   0.1 &  0.0 &  0.0 &  0.0 \\
                          & net-zero 2050 &   0.2 &  0.0 &  0.0 &  0.0 \\
15, Geothermal & Delayed transition &    NA &  0.0 &  0.0 &  0.0 \\
                          & NDCs &   0.0 &  0.0 &  0.0 &  0.0 \\
                          & net-zero 2050 &   0.1 &  0.0 &  0.0 &  0.0 \\
17, Conventional hydropower & Delayed transition &    NA &  0.0 &  0.0 &  0.0 \\
                          & NDCs &   0.1 &  0.0 &  0.0 &  0.0 \\
                          & net-zero 2050 &   0.1 &  0.0 &  0.0 &  0.0 \\
20, Wind & Delayed transition &    NA &  0.0 &  0.0 &  0.0 \\
                          & NDCs &   0.0 &  0.0 &  0.0 &  0.0 \\
                          & net-zero 2050 &   0.1 &  0.0 &  0.0 &  0.0 \\
24, Solar photovoltaic (PV) with tracking & Delayed transition &    NA &  0.0 &  0.0 &  0.0 \\
                          & NDCs &   0.1 &  0.0 &  0.0 &  0.0 \\
                          & net-zero 2050 &   0.3 &  0.1 &  0.0 &  0.0 \\
25, Solar PV with storage & Delayed transition &    NA &  0.0 &  0.0 &  0.0 \\
                          & NDCs &   0.1 &  0.0 &  0.0 &  0.0 \\
                          & net-zero 2050 &   0.3 &  0.1 &  0.0 &  0.0 \\
\bottomrule
\end{tabular}
		\caption{MT/year of NET capacity required for financial net-zero with respect to carbon emissions over the length of the financing. 
			NA indicates that there is no solution for the financing maturity considered.  This happens with Delayed transition because the carbon price is zero for the first 10 years so we assume no profit from selling carbon offsets is possible.  Zeros indicate that the required NET capacity is below 0.1MT/year.
		}
		\label{t:fnz}
		\end{center}
	\end{adjustwidth}
\end{table}

Table \ref{t:spread} gives the annuity spreads required to pay for the carbon emissions.  Each technology has a 4 by 4 array of results: three NGFS carbon price scenarios, one NET price scenario assuming that the project incorporates a NET facility, and four financing scenarios, 10, 20, 30 years, and to end of operation.   If the end of operation is shorter than the financing period, then the financing period is to the end of operation.

Non-renewable technologies have the general pattern of increasing annuity spreads for increasing financing length.  This is the opposite of usual expectations where if payments are spread over a longer period, then they become lower.  The reason for this increase is that carbon prices increase monotonically in all NGFS price scenarios.  

Renewable technologies also have a general pattern, and it is opposite to the non-renewable, i.e., decreasing annuity spreads with increasing financing length.  This is because the only significant carbon emissions are during construction.

Table \ref{t:fnz} shows the required size of an accompanying  NET project to make each power project financially net-zero.  That is, the project builds the power plant, and sufficient NET capacity that it can sell so that the net carbon cost is zero.  These NET capacity requirements are of similar size to the carbon emissions of the original projects, as would be expected.  Where such combined power and NET projects are feasible, then they are the most attractive solution from a financial point of view.

\section{Conclusions}

We introduce the Carbon Equivalence Principle (CEP).  The CEP states that the volume and timing of carbon flows enabled by financial products shall be booked as  \xca termsheets linked to the original product termsheets.  The first implication of this principle, is that all financial products are already ESG-linked products.  Note that the CEP does not turn financial products into ESG-linked products artificially, instead it reveals where they are already ESG-linked.

We illustrate the CEP in action by looking at project finance for  new electricity generation, and immediately see its usefulness.  The whole structure of the financing and project calculation is changed: the project may become a stranded asset, thus changing the recovery rate for the financing; the project may only be viable if a second NET project is financed in parallel; NET technologies need to be developed so that they are mature, and there is the possibility that projects can be financial net-zero. 

To get to net-zero equivalent carbon emissions on loan portfolios, banks require visibility on the associated carbon flows of the projects so as to enable accurate accounting using standard bank systems.  Application of the CEP provides this. 

Finally, simply making carbon flows visible in time and volume can fundamentally change the pricing and design of financial products.  From a financial point of view, the target is financial net-zero and this requires carbon net-negative.

\FloatBarrier
\section{Appendix}

\subsection{Financial calculation setup}

The financial setup for the numerical results in Tables \ref{t:spread}, \ref{t:fnz}, and Figure \ref{f:ngfs} is summarized below:
\begin{description}
	\item[As-of:] This indicates the calculation date, and project start date, 2021-11-30
	\item[Electricity generation projects:] References \cite{us2020capital,us2021capital} provide details on cost, engineering, and operating emissions, these are combined into Table \ref{t:technology}.  Construction emissions are taken from \url{https://data.nrel.gov/submissions/171}.  Capacity factors are taken from \url{openei.org}.
	\begin{itemize}
		\item Each project has four phases: plan; build; operate; deconstruct.  We assume no carbon is emitted in the planning phase.
		\item Deconstruction is assumed to take half the build time and 5\%\ of the build cost.  Results are not materially affected by variations of these assumptions.
	\end{itemize}
	\item[Negative emissions projects:] Carbon Capture and Storage (CCS) can be built into powerplants.  This can be efficient for gas and coal plants because the CCS will then have a higher-than-atmosphere percentage of carbon dioxide to start from.  However, this typically leaves 10\%\ or more of the carbon not captured.  Thus it can  be advantageous for a powerplant to build a Negative Emissions Technology plant at the same time.  Market carbon prices can be far above the NET costs because of limited NET capacity.  Price differences because of capacity limitations are commonly observed, for example, in gas markets where prices are seasonable because of the lack of storage capacity.  This is also why pairing a NET project and a powerplant is useful if it is feasible.
	\begin{itemize}
	\item Costs in 2011 USD for NET are shown in Table \ref{t:net} following \cite{fuss2018negative}.
	\item Consistent with usual IAMs we assume NET at utility scale are only available from 2030, except for afforestation and reforestation which are avavilable now.  
	\item Cost multipliers for emerging NET at utility scale in 2030 are six times for the lower bound, and three times for the upper bound,following data in \cite{fuss2018negative}.  These decrease linearly to 2050 when we assume the technologies are mature, so costs are constant in constant dollars thereafter and then adjusted for inflation.
	\end{itemize}
	\item[Carbon price scenarios:]  NGFS scenarios from \url{https://data.ene.iiasa.ac.at/ngfs}.  
	\begin{itemize}
		\item We use three scenarios: net-zero 2050; Nationally Defined Constributions (NDC); and Delayed Transition, which means zero price for cabon for the first ten years.  
		\item 2010 USD are converted to USD for each future date by adjusting for inflation using historic US CPI index values and US CPI swaps reported at FRED \url{https://fred.stlouisfed.org/}.
	\end{itemize}
	\item[Financing:] We consider four lengths for financing: 10 years; 20 years; 30 years; and operating lifetime. We always take the minimum with the operating lifetime.
	\item[Financial market:] We use a US setting with market data from the US Federal Reserve, St. Louis, FRED \url{https://fred.stlouisfed.org/}.
	\begin{itemize}
		\item As-of date for market data: "as-of" terminology, see above
		\item Riskless discount rates: derived from SOFR swaps
		\item Risky discount rates using spread between SOFR discount rate and BBB discount curve at 5y, this spread is observed to be 157bps.  We use a flat spread between riskless and risky.
	\end{itemize}
	\item[Sanity check:] We take an alternative source of carbon emissions \url{https://www.coaltrans.com/insights/article/ecocarbon-august-2021} to perform a sanity check on the level of the results.  This is the simple calculation shown in Table \ref{t:check}.  The result is similar to Case No. 01, from Table \ref{t:technology}, which is also for a coal-fired power plan.
\end{description}

The setup for electricity generation technologies  is given in Table \ref{t:technology}. The mid-price of the cheapest NET of Table \ref{t:net} is shown in Figure \ref{f:ngfs} on the left, and on the right an inflation adjusted version of the NET costs.  The cheapest NET is afforestation and reforestation, when it is available.
\begin{table}
	\begin{adjustwidth}{-3.25cm}{-1cm}
		\begin{tabular}{lp{1cm}p{1cm}p{1.4cm}p{1cm}p{1cm}p{1cm}p{1cm}p{1cm}}
			\toprule
			Case No., Technology &  Size (MW) &  Capital cost (BUSD) &  Develop (months) &  Build (months) &  Lifespan (years) &  Capacity Factor  &  Carbon per year (MT) &  Carbon to build (MT) \\
			\midrule
			01, Ultra-supercritical coal (USC) &        650 &                2.552 &                    24 &              36 &                40 &                         0.850 &                  3.91 &                      1.37 \\
			03, USC with 90\% CCS &        650 &                4.079 &                    24 &              36 &                40 &                         0.850 &                  0.57 &                      1.37 \\
			08, Combined-cycle—single shaft &        418 &                0.484 &                    18 &              24 &                40 &                         0.870 &                  1.09 &                      0.12 \\
			09, Combined-cycle with 90\% CCS &        377 &                0.999 &                    24 &              30 &                40 &                         0.870 &                  0.11 &                      0.11 \\
			12, Nuclear—small modular reactor &        600 &                3.967 &                    24 &              48 &                40 &                         0.900 &                  0.00 &                      0.42 \\        
			15, Geothermal &         50 &                0.135 &                    24 &              36 &                40 &                         0.950 &                  0.00 &                      0.26 \\
			17, Conventional hydropower &        100 &                0.568 &                    36 &              36 &                50 &                         0.500 &                  0.00 &                      0.27 \\
			20, Wind  &        200 &                0.270 &                    12 &               9 &                25 &                         0.380 &                  0.00 &                      0.48 \\
			24, Solar photovoltaic (PV) with tracking  &        150 &                0.210 &                    12 &               6 &                30 &                         0.158 &                  0.00 &                      1.70 \\
			\bottomrule
		\end{tabular}
		\caption{Technology characteristics, financial, timing, and carbon emissions.  Most columns derived from US Energy Information Administration, see text for more details.  MT = mega Tonne, BUSD = US Billion Dollars, MW = mega Watt.}
		\label{t:technology}
	\end{adjustwidth}
\end{table}

\begin{table}[]
	\begin{adjustwidth}{-2.25cm}{-1cm}
\begin{center}
\begin{tabular}{lrrr}
\toprule
Negative emissions   technology (NET) & \multicolumn{1}{r}{Mature from year} & \multicolumn{1}{r}{2011\$/Tonne low} & \multicolumn{1}{r}{2011\$/Tonne high}  \\
\midrule
Forest       & 2021                                    & 5                                    & 50                                \\
BECCS                                 & 2050                                    & 100                                  & 200                                     \\
DACCS                                 & 2050                                    & 100                                  & 300                                     \\
Enhanced weathering                   & 2050                                    & 50                                   & 200                               \\
Biochar                               & 2050                                    & 30                                   & 120                                       \\
Soil carbon                           & 2050                                    & 0                                    & 100                                      \\
\bottomrule
\end{tabular}
\end{center}
\caption{Negative emissions technologies costs for utility scale sequestration plants, from present if available now, or from 2050 when assumed mature, see text for details.  We use an emergent technology multiplier on costs from 2030 to 2050.}
\label{t:net}
\end{adjustwidth}
\end{table}

\subsection{Sanity check on carbon emissions volume}

We start with the sanity check shown in Table \ref{t:check}.  This calculation is similar to the carbon emissions calculation for each financing case and technology, and it uses a separate data source.  We see that a 650MW coal-fired power plant can emit around 4MT of carbon per year in operation with a capacity factor of 0.85.  This already gives an orientation for the further results: if carbon prices are USD100 per year, this adds nearly half a billion dollars yearly cost for a capital outlay of around USD 2.5B in Case No. 01, Ultra-supercritical coal.  USD100 per Tonne of carbon is reached in many NGFS scenarios within 10 years.  

\begin{table}[h]
	\begin{center}
		\includegraphics[trim=50 660 90 0, clip, width=1\textwidth]{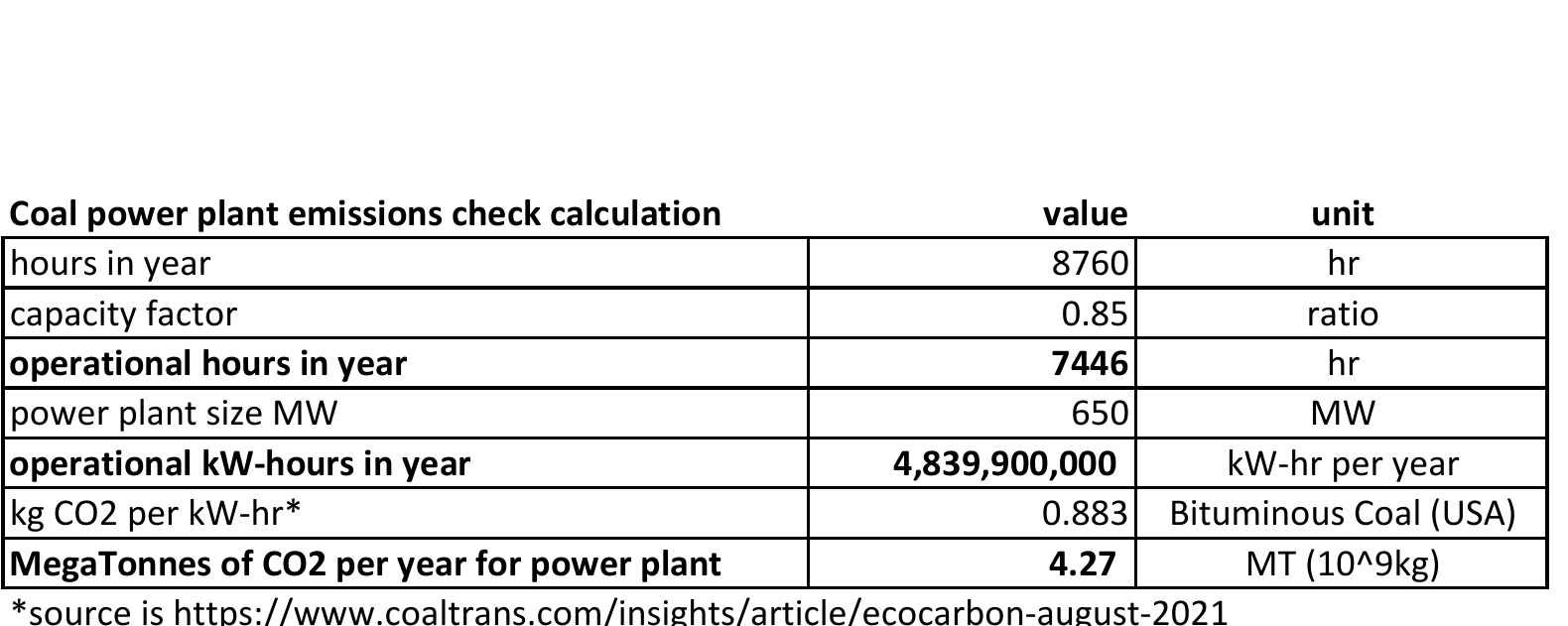}
	\end{center}
	\caption{Sanity check on level of yearly carbon emissions from a coal-fired power plant using alternative data source. Similar yearly volume as Case No. 01, 4.27MT versus 3.91MT, so sanity check passes.}
	\label{t:check}
\end{table}

\subsection{Financial calculations}

\paragraph{Carbon costs as annuity spread.}
The price of an annuity with notional $N$, maturity $T$, using discount curve $r$ is standard \cite{brigo2006interest}\ :
\begin{equation}
\text{NPV}_\text{annuity}(T,r) = N \times \sum_{i=1}^{i=n}  D_r(t_i) dc_\alpha(t_i, t_{i-1})
\end{equation}
where the annuity has parameters: $n$ payments at pay dates $t_i$ up to maturity $T=t_n$; uses day counter $\alpha$; and discount factors $D_r(t_i)$.  The day counter $dc_\alpha$ is the calculation method from pairs of dates to fractions of a year.  Typical ones in use are Actual/Actual, 30/360, and official definitions can be found on the ISDA website \url{https://www.isda.org}.  We use Actual/Actual.  We ignore date rolling conventions, and date adjustments as non-material to this analysis.  These change pay dates, so payments are always on business days, and may also adjust how the accrual period is calculated.  The accrual period is the period between unadjusted payment dates.  The discount factors $D_r$ use the discounting curve $r$.  Here we take $r$ as the riskless rate, which we take by convention to be defined by the SOFR swap curve, see \url{https://www.newyorkfed.org/markets/reference-rates/sofr} for details.  Details of curve construction do not materially affect the results.  

Ignoring counterparty risk, the value of a non-negative spread $s$ on an annuity is just $s \times\text{NPV}_\text{annuity}(T,r)$.

\paragraph{Financing cost change in case of stranded assets.}  Financing cost expresses the sum of the cost of the bank's funding and the expected loss from loan default.  For simplicity we assume that the bank can fund at the riskless rate, i.e., SOFR.  We also assume that the projects are designed to have a BBB credit rating, either by ratings agencies, or by bank-internal credit departments.  This is the typical tradeoff between guaranteed wholesale electricity purchase price, and financing costs for the project.  BBB is the lowest investment grade rating.  

The situation is more complex than it appears because the recovery rate observed for power infrastructure to the time of writing \cite{sandp2021recovery} may not be achieved.  If  repayment is pushed out the carbon prices will be higher, potentially requiring repayment to be pushed out further, etc.  In fact, the power plan may become a stranded asset at that point.  Thus we must adjust the recovery rate to take the possibility of becoming a stranded asset into account, increasing the financing costs assuming that the project has the original BBB rating.  Note that we assume the credit rating describes the default probability, and not the expected loss.

For senior unsecured debt, the recovery rate on power plants historically is 69\%\ +4.5\%\ = 73.5\%, where the 4.5\%\ adjusts for restructuring, not bankruptcy \cite{sandp2021recovery}.   Prior to carbon pricing, power plants typically restructured to pay over a longer period, rather than going bankrupt.  With carbon pricing, if the cost is high enough then we can expect a much lower recovery rate, possibly even close to zero if the technology has become non-competitive.  That is, the power plant itself becomes a stranded asset.

\paragraph{Financial net-zero.}  We want to get enough NET capacity so that 1) we can make the NET capture permanent, and  2) there is sufficient profit  from permanent NET capacity sales on the regulatory market to make a profit to pay for emissions over the life of the financing.  Thus we solve for a constant capacity that creates sufficient profits to pay for the annuity spread of the carbon emissions.  This achieves financial net-zero for the project financing.  That is, the NET capacity $v_\text{NET}$ solves:
\begin{equation}
v_\text{NET} \times \int_{t=0}^{t=T} \max(0, p(t)_\text{NGFS scenario} - c(t)_\text{NET}) = \text{NPV}_\text{carbon emissions}
\end{equation}
where $p_\text{NGFS scenario}$ is the price for one unit of carbon-equivalent emissions permits, and $c_\text{NET}$ is the unit cost for creating permanent NET capacity, and $\text{NPV}_\text{carbon emissions}$ is the cost of the carbon emissions of the project.  Here we define permanent as 100 years from the date of sequestration, and  $T$ is the maturity of the financing to the combined project, power and NET.

\section{Acknowledgments and Disclaimer}

The authors would like to gratefully acknowledge discussions with Imane Bakkar, Susumu Higaki, Cathryn Kelly, Seiya Tateiri, Niels Van Vliet, and Kohei Ueda.

\section*{Author Contributions}

All authors contributed equally to the paper.

\section*{Data availability}

Data is available on request.

\section*{Code availability}

Code is available on request.

\section*{Competing interests}

The authors declare the following competing interests: CK is also global head of quant innovation at MUFG Bank, and global head of XVA quant research at the same institution; M.B. is also head of XVA and counterparty credit risk modeling at Lloyds Banking Group, and non-executive director at LCH.

{\bf Disclaimer}.  The views expressed by the authors C.K. and M.B. in this paper are personal and do not necessarily reflect the views or policies of current or previous employers. No guaranteed fit for any purpose. Use at your own risk.

\bibliographystyle{chicago}
\bibliography{manifesto3}
\end{document}